\documentclass{elsart}

\usepackage{graphicx}

\begin{document}
\begin{frontmatter}

\title{Network structure and dynamics of hydrogenated amorphous silicon}


\author{D. A. Drabold, T. A. Abtew, F. Inam and Y. Pan}

\address{Department of Physics and Astronomy, Ohio University, Athens, Ohio 45701-2979, USA}

\begin{abstract}
In this paper we discuss the application of current {\it ab initio} computer simulation techniques to hydrogenated amorphous silicon (a-Si:H). We begin by discussing thermal fluctuation in the number of coordination defects in the material, and its temperature dependence. We connect this to the ``fluctuating bond center detachment" mechanism for liberating H bonded to Si atoms. Next, from extended thermal MD simulation, we illustrate various mechanisms of H motion. The dynamics of the lattice is then linked to the electrons, and we point out that the squared electron-lattice coupling (and the thermally-induced mean square variation in electron energy eigenvalues) is robustly proportional to the localization of the conjugate state, if localization is measured with inverse participation ratio. Finally we discuss the Staebler-Wronski effect using these methods, and argue that a sophisticated local heating picture (based upon reasonable calculations of the electron-lattice coupling and molecular dynamic simulation) explains significant aspects of the phenomenon.
\end{abstract}

\ead{drabold@ohio.edu}
\end{frontmatter}

\section{Introduction}
\label{I}
The initial challenge in modeling a disordered material is to answer the question: `where are the atoms?'. Even for static properties this is a challenge that persists, especially in multinary systems (for example the `phase change' chalcogenides and/or disordered systems with nearly degenerate conformations). For materials that form glasses, molecular dynamics simulation of quenching from a liquid usually will produce reasonable computer models \cite{cobb95}\cite{tafen05}. For systems that do not form glasses (like a-Si or a-Ge), such melt quenching is usually unsatisfactory, producing too many defects\cite{dadcurrentopinion}.

For a-Si, an accepted structural model exists: the continuous random network, usually constructed from the recipe of Wooten, Weaire and Winer (WWW) or improved variants\cite{www,dtw}. When hydrogen is added to model a-Si:H (the material of real practical interest), the situation is less clear, but H is believed from experiments to passivate defects (especially undercoordinated `dangling bond' sites),  attach to bondcenter or other sites \cite{santos92}\cite{chris95}\cite{Lanzavecchia96}\cite{biswas98}\cite{franz98}, and in some cases exists in molecular form \cite{fedders00}. NMR provides evidence that microvoids sometimes contain H\cite{dick}. 

In this paper, we briefly discuss the structure of a-Si:H, and focus primarily on the network or lattice dynamics of the system, both in the electronic ground state and in an electronic excited state, the latter to gain insight into light-induced phenomena in the material. We point out special features of the atomic dynamics that lead to processes unique to disordered systems. We indicate a path to modeling these using current \emph{ab initio} simulation techniques, and show how such simulations can be helpful to interpreting observations of light-induced changes. The dynamics of the material, and especially its coupling to the electron states, is important for IR imaging and detection devices based upon a-Si:H.
The paper is organized as follows. Section \ref{II} describes the methods used to compute the results of the paper. Section \ref{III} describes certain stable dihydride conformations. Section \ref{IV} discusses network dynamics of a-Si. In Sec. \ref{V} we focus more specifically on the motion of H in such a network, and discuss a new mechanism for ``stripping" chemically bonded H from the network, and show that this is a necessary to understand H diffusion in the network. In Sec. \ref{VI}, we make the link between the electronic and lattice system clearer, and show that the electron-phonon coupling is always large for localized states, and in Sec. \ref{VII} we discuss our simulations of light-induced phenomena in terms of a local heating picture based upon {\it ab initio} interactions.

\section{Constraints on Theory}
\label{II}
In this paper we report results using the code SIESTA \cite{soler02}, a local basis density functional code suited to studies of a-Si:H. Most of the simulations discussed are in the electronic ground state; for photo-excited simulations, charge states are modified to model photo-induced occupation changes of the single-particle levels. Prior work \cite{chris94}\cite{atta04} has shown that the energetics of H in Si is delicate, even in the crystal, and especially dependent upon the basis set used to represent the valence electrons. For this reason, a double-zeta polarized basis is always used for H and a minimum of double zeta for the Si.

\section{Network statics: dihydride conformations in an amorphous matrix}
\label{III}
\begin{figure}
\center
\resizebox{90mm}{!}{\includegraphics{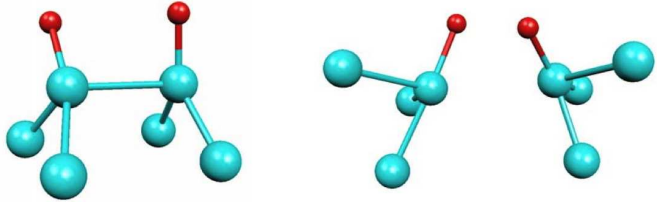}}
  \caption{Two dihydride configurations. Left: type ``A", Right, type ``B". In the amorphous matrix, type A exhibits proton separations near 2.2~\AA and type B near 1.9~\AA~.}
\label{fig1}  
\end{figure}

There are many preferred positions for atomic H in an a-Si matrix. These have been extensively discussed elsewhere \cite{abtew06}\cite{abtew006}. In this section, we mention only the case of dihydrides, both because they are important in some theories of the SWE, and also because we find that they emerge naturally in our simulations of a-Si:H.  Taylor and coworkers\cite{su02} have used proton NMR to measure the distribution of proton separations in various samples of a-Si:H. Early work suggested the importance of $2.4$~\AA~ \cite{su02} as a preferred proton-proton separation; more recent work emphasizes distances closer to $1.8$~\AA~\cite{bobela}. The latter work has caused us to consider additional dihydrides (see Fig. \ref{fig1}). The procedure is to build desired dihydride structures ``by hand" within the network (eg in the solid state, not molecules), and then relax the positions of all the atoms to see the final separation between protons. We have previously reported \cite{abtew05} in detail on proton separation in SiH$_2$ (two H on one Si), consistently obtaining a distance of about $2.4$~\AA~.  For the topologies of Fig. \ref{fig1}, we tried three different starting configurations for each type and found final mean proton separations of $2.19$~\AA~ for type A, and $1.88$~\AA~ for type B. As for the case of SiH$_{2}$, the relaxed final conformations all had very similar proton spacings, despite quite different starting configurations. We have reported time-averaged distributions of proton separations in the solid state\cite{abtew06}.
\section{Network dynamics: coordination fluctuation in a-Si}
\label{IV}
For crystalline systems, atomic motions are constrained by the regularity of the lattice.  In an amorphous system there may be significant variations in the coordination at moderate temperature\cite{tunafin}\cite{dadbook}. Such coordination fluctuations have not been directly observed experimentally, though the disorder clearly causes significantly different dynamics in the amorphous phase. The point is also evident from the large thermal broadening effects on the pair-distribution function, treated in the harmonic approximation by Feldman and coworkers\cite{feldman} and by the path-integral Monte Carlo studies of Herrero\cite{herrero}.

\begin{figure}
\center
\resizebox{100mm}{!}{\includegraphics{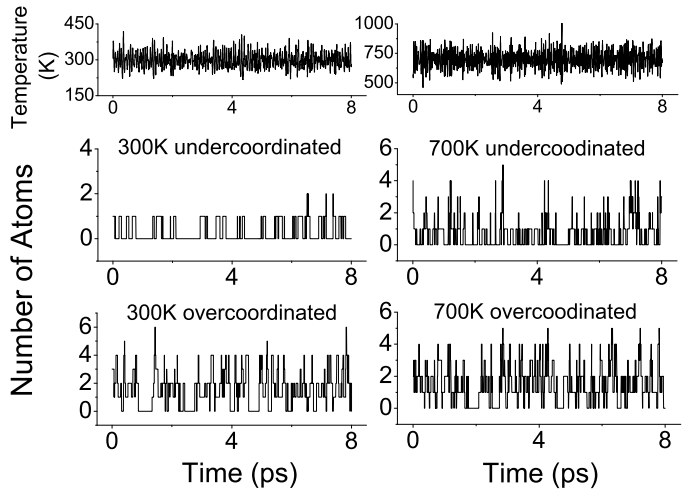}}
  \caption{Thermally-induced coordinate fluctuations in 64-atom a-Si mode, T=300K and T=700K shown. The coordination radius is $r_{c}=2.75$~\AA~ (see text).}
\label{fig2}
\end{figure}

To illustrate the variation in instantaneous coordination, we have performed 8 ps thermal simulations of a 216 atom WWW-type model of a-Si:H at 300K and 700K.  A Nos\'e  thermostat\cite{nose} was used to maintain the desired temperatures, and the volume was held fixed. An initial 1 ps simulation was performed beginning with a relaxed configuration before accumulating data on coordination fluctuation, to equilibrate the system to the desired temperature.  To quantify these fluctuations we to define a coordination distance $r_{c}$. Evidently the fluctuations resulting from the dynamics are sensitive to $r_{c}$. In Fig. \ref{fig2}, we have selected r$_c=$2.75~\AA. This is close to the minimum after the first-neighbor peak. While the details of the numbers of improperly coordinated atoms depends upon $r_{c}$, the qualitative appearance of a random telegraph signal is similar for all $r_{c}$. 


A na\"ive reading of Fig. \ref{fig2} would suggest that electronic gap states should be created and destroyed on fs time scales if there was a one-to-one equivalence of coordination and electronic or "spectral" defects. While this cannot be the case\cite{tunafin}, it is true that the electronic eigenvalues near the gap are very strongly modulated by the network dynamics. It is clear that this must be important to carrier transport, since the Fermi level is near these strongly modulated states. We connect this to the Meyer-Neldel relation in another paper in this volume\cite{mnrmz}.

\begin{figure}
\center
\resizebox{100mm}{!}{\includegraphics{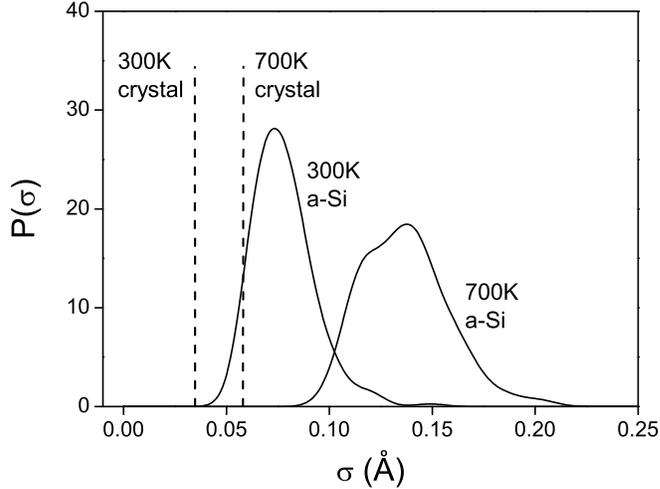}}
  \caption{Distribution of root-mean-square fluctuations $\sigma$ in the ionic positions in 216-atom model at T=300K and T=700K, for both diamond and a-Si and constant T simulation, using empirical potential\cite{edip}. The data was acquired over 200 ps.}
\label{fig3}
\end{figure}

In Fig. \ref{fig3}, we quantify the equilibrium distributions of root-mean-square variation about equilibrium positions for both amorphous and
crystalline Si. The means of these distributions are slightly smaller than those reported by Herrero\cite{herrero}, possibly because the latter include (atomic) quantum delocalization effects. The broad distributions for the amorphous phases convey new information about the variability of the local structural fluctuations of the amorphous network unseen in crystals. Note particularly the development of a ``tail" for a-Si at 700K. Here we used the empirical Environment Dependent Empirical Potential ``EDIP"\cite{edip}. 

\section{Hydrogen motion}
\label{V}
The dynamics of H in a-Si:H is one of the key areas of basic research in the material. As is well known, H plays a critical role in passivating dangling bonds on undercoordinated atoms, but is also deeply connected to the light-induced metastability of the material \cite{fedders92}\cite{jackson88}\cite{branz99}\cite{abtew006}, the Staebler-Wronski Effect (SWE)\cite{fritzsche01} that limits the utility of the material for use as a photovoltaic.

From extended thermal simulation using accurate interactions, we have detected an important mechanism for liberating chemically bonded H in a-Si:H \cite{abtew07}. The observation is that the finite-temperature amorphous matrix allows short-lived structural fluctuations such that a H atom initially bonded to a Si may convert to an instantaneous bond-center (BC) configuration when an additional Si is near enough to the H. When in a BC conformation, the H is much more likely to escape than if it is bonded to the original Si. What this work reveals is a novel aspect of the short time dynamics of the H, and a mechanism that could never occur in a crystal because of the constrained dynamics. Other important mechanisms exist (for example BC to BC hopping), but these occur on longer time scales that are difficult to treat with MD simulation (while clearly important to the physical properties of the material). Fluctuating Bond-Center Detachment (FBCD)\cite{abtew07} is the primary means to convert bonded H to `free' H able to hop, and is therefore an important ingredient to understanding H motion. FBCD is possible precisely because of the fluctuations of the preceding section.  FBCD is also important to electronic properties, since it is creates dangling bonds. Our mechanism is reminiscent of the work of Su and Pantiledes\cite{suu02}, though considerably more general and different in details.  To check our work, we have run these simulations with a plane wave code (VASP) and found similar results\cite{vasp}. Illustrations and details of FBCD are available elsewhere\cite{abtew07}.

FBCD is primarily relevant as a means for creating free atomic H, which has a short lifetime, quickly being absorbed by a BC or possibly to passivate  dangling bond. This shows that dangling bonds are strongly preferred destinations for mobile H. On longer time scales, there is certainly hopping between BC and other sites. This has not yet been thoroughly explored, though \emph{ab initio} calculations have provided explicit paths or such events. Such events are illustrated elsewhere\cite{dadbook}.

\section{Electron-lattice coupling}
\label{VI}
\begin{figure}
\center
\resizebox{90mm}{!}{\includegraphics{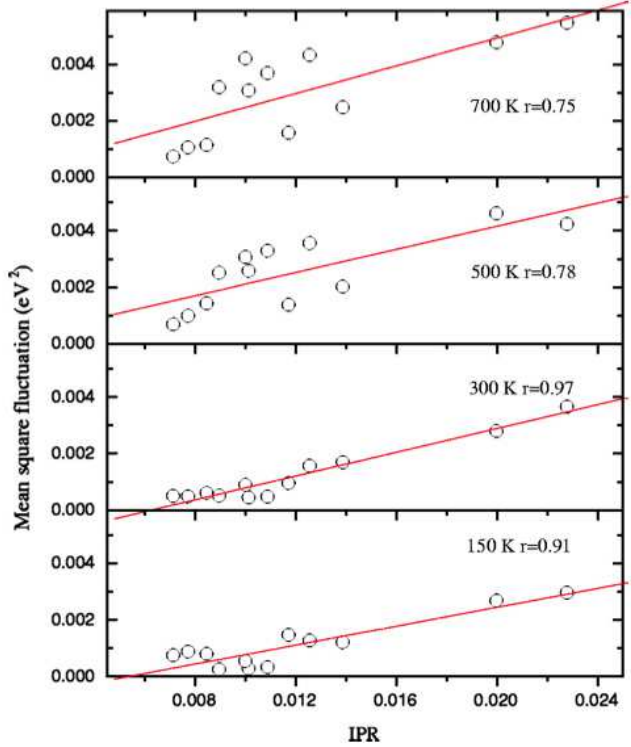}}
\caption{Variance of eigenvalues near the gap plotted against inverse participation ratio (IPR) at various temperatures. The fluctuations at T=150K and 300K are found to be linearly correlated with the IPR, The correlation coefficient for the linear fit is given. Simulation was 2.5ps on 216 atom cell with Nos\'e dynamics (from \cite{fynn04}).}
\label{fig4}
\end{figure}

To understand either carrier transport or light-induced effects, the coupling between electrons and phonons is the key. In any amorphous semiconductor, the region of the optical gap consists of highly localized defect states near the middle of the gap (dangling bonds in a-Si:H, depending somewhat upon charge state), and "tail" states, which arise from distortions of various kinds in the network. We have shown that there is a strong correlation between the degree of localization (spatial confinement of electron states), and how sensitive these electrons (more specifically, their energies) are to lattice motions \cite{fynn04}\cite{tunafin}. If one measures the localization by the inverse participation ratio (IPR)\cite{ziman}, it is possible to show that for highly localized states, the electron-phonon coupling is directly proportional to the IPR. Thus, if $\langle \delta\lambda_{n}^{2} \rangle$ is the thermally-induced variation in electron energy eigenvalue $\lambda_{n}$ and $I_{n}$ is the inverse participation ratio of state $\Psi_{n}$ conjugate to $\lambda_{n}$ and $\Xi_{n}(\omega)$ is the electron-lattice coupling between electron $n$ and phonon  $\omega$, then: (1) $\langle \delta\lambda_{n}^{2} \rangle \propto I_{n}$ and (2) $\Xi_{n}^2(\omega) \propto I_{n}$\cite{gamma}. These simple relations can be derived with some crude assumptions, but are also borne out quantitatively in thermal MD simulations. As the localization ($I$)  is a very strong function of the energy (large near midgap, decaying through the band tails and reaching a constant low value well past the mobility edge), it means that the response of electrons with energies in different parts of the gap will experience quite different response to the lattice. This linearity is revealed in Figure \ref{fig4}. The implications to transport are discussed elsewhere\cite{transport}.

\section{Staebler-Wronski effect}
\label{VII}
\subsection{Approach}
The preceding discussion is a natural preliminary to modeling the SWE.  The approach we outline here is an atomistic one, and has ideas in common with phenomenological theories, though the exact connections are sometimes difficult to determine, in part because of issues like the difference in time scales between simulations and experiments. The novel features of our studies are (1) the use detailed microscopic information in modeling SWE (realistic structural models of a-Si:H) and (2) employing current techniques capable of explicitly handing the details of localization due to disorder and defects, and from which interatomic forces in both the electronic ground state and excited states may be reasonably obtained.\\

In any amorphous insulator there are defect states near the middle of the optical gap, tail states that extend into the gap; these tails are exponential and their rate of decay into the gap is temperature dependent. A na\"ive view of the interaction of light with a solid is that a photon may promote an electron from a valence state to a conduction state in such a way that energy is conserved in the process.\\

Consider a-Si:H with the Fermi level near the middle of the gap. Then states near the Fermi level are necessarily somewhat localized (highly localized if near the gap center). When one computes the forces on atoms within any \emph{ab initio} approach, the electronic contribution (the so-called bandstructure force) is obtained  as: $F_{\alpha}=-2\sum_{n}\langle \Psi_{n}|\partial H/\partial R_{\alpha}|\Psi_{n} \rangle$  . Here, $H$ is the single-particle (usually Kohn-Sham) Hamiltonian, $\Psi_{n}$ is an exact eigenvector of $H$, $R_{\alpha}$ is the position of atom $\alpha$ , $2$ is for spin, and the index $n$ runs over all occupied states. Within such a picture, it is natural to model light-induced occupation changes by modifying the set of occupied and unoccupied states. The logic is that if the change in occupation (or charge state) causes a significant change in the force, it may be sufficient to cause significant rearrangements in the network, perhaps even the formation of new defects. We have used this approach in a-Se \cite{zhang99}, a-As$_2$Se$_3$\cite{li00} and a-Si \cite{fedders92}. This approach is ideal for checking the structural/dynamical consequences of modifying the charge state of a well-localized dangling bond.\\

We have made occupation changes of the type discussed here on a variety of models, and reported the results elsewhere. In a nutshell, it was found that (1) occupation changes in dangling bonds often lead to creation of new defects, and in particular, new dangling bonds; (2) in several simulations we observed that dihydride conformations were created; (3) enhanced H motion for the simulated photoexcited state.

\subsection{Local heating}
The picture we adopt is that the change in Hellmann-Feynman forces associated with the occupation changes can modify the network and we have evidence that the resulting modifications are not inconsistent with experiment (defect creation, and even a tendency to create dihydride sites when we perform excited state simulations). This picture is familiar: it is a variant of a local heating model.  For example, if an electron is promoted to a low-energy conduction or deep state near the conduction edge $\Psi$, then an additional force is added to all the atoms, and has the form $F_{\alpha}=\langle \Psi|-\partial H/\partial R_{\alpha}|\Psi \rangle$ . The interesting feature of this equation is that $F_{\alpha}$ is highly nonuniform in the cell. In particular, if $\Psi$ is localized, then only the volume of space in which $\Psi$ has support contributes to the force shift. Moreover, it is typical that such states have most of their weight on a few sites and the remainder spread out over many other sites. As such, the atoms contributing the largest fraction to the localized charge will experience the largest forces and the response of the system to occupation is likely to be most dramatic near these sites. This is conceptually the same as the "hot spots" proposed by Biswas and coworkers long ago \cite{biswas91}, though we believe that a classical simulation is unlikely to ever succeed quantitatively because such an approach cannot include the electron-lattice interaction properly. There are additional factors to be considered of course: the local rigidity of the network may or may not allow large rearrangements. Yet, the very fact that the state was localized in a particular region in the first place implies that the local geometry and/or chemistry was very different from the "mean" and thus, in all likelihood, mechanically and vibrationally distinct from the rest of the material. Note that if $\Psi$ is extended, then there will be contributions for all atoms, but will be small, and would create no structural changes - so our approach produces no rearrangements in crystals with its extended (Bloch) states.

\begin{figure}
\center
\resizebox{90mm}{!}{\includegraphics{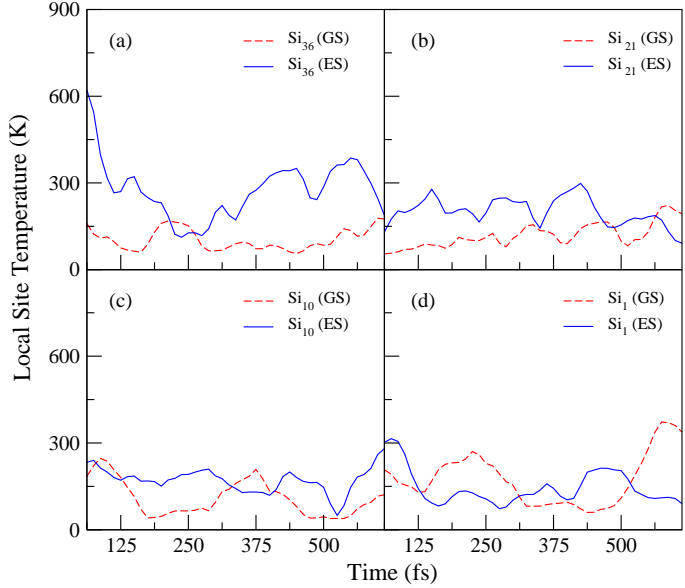}}
\caption{Local temperature (kinetic energy)  at dangling bond site (Si$_{36}$) and other sites unconnected with the localized state centered on Si$_{36}$. GS means electronic ground state, ES, excited state.}
\label{fig5}
\end{figure}

To gauge the validity of this scheme, we have performed a new simulation using our usual approximations (71 atom cell, SIESTA, double-zeta (Si) double-zeta polarized (H) basis set, 10$^4$ steps at 0.25fs/step at T=300K, globally maintained by a Nos\'e thermostat). Consistent with prior simulations we found that (1) in the electronic ground state, no dihydride species were formed; (2) with an electronic promotion (involving change of occupation of a dangling bond state) a SiH$_{2}$ was formed. We also tracked the local temperature in the vicinity of the dangling bond suffering occupation change. In Figure \ref{fig5} we plot the local temperature at the dangling bond site and other sites not associated with the localized state. The excited defect thermalizes in less than a quarter ps through the cell. The newly created SiH$_2$ was formed near the defect-induced ``hotspot".

We have observed that dihydrides tend to form in our photoexcited simulations. There are two possible explanations for this. The first is that there is something special about the photoexcited state that favors the creation of dihydride structures, a hypothetical chemical effect that increases the likelihood of dihydride formation. It is difficult to see why this should be the case, and our preference is for a second, local heating picture. We note that dihydride formation is extremely rare or nonexistent on picosecond time scales at room temperature. On the other hand, ground state simulations at higher temperatures do sometimes reveal H hopping that leads to the creation of dihydrides. The simplest explanation is that if a localized state undergoes an occupation change, and the local structure is suited to creating a dihydride (for example if there is a pre-exisiting monohydride, and free H available), then the dihydride may indeed form. Implicit to this argument is a belief that the energy of dihydride sites  is favored over monohydrides. While this may be true, we have not been able to prove it. There is a good deal more to be done on this approach, but hope that it will contribute significantly to understanding of light induced metastability for a-Si:H and other materials.

\section{Conclusions}
\label{VIII}
We have presented recent work using first principles methods to study the atomistic of hydrogenated amorphous silicon. We have dwelled upon the role of H in the material; both its statics and dynamics. Generic features of the lattice-electron coupling are deduced for localized states, and we make use of this to approximately, but directly simulate the effects of photoexcitation.

\section*{Acknowledgements}
We have benefited from discussions with P. A. Fedders, P. C. Taylor and E. A. Schiff. Research was sponsored by the U.S. Army Research Office and U.S. Army Research Laboratory and was accomplished under Cooperative Agreement Number W911NF-0-2-0026.  The views and conclusions contained in this document are those of the authors and should not be interpreted as representing the official policies, either expressed or implied, of the Army Research Office, Army Research Laboratory, or the U.S. Government.  The U.S. Government is authorized to reproduce and distribute reprints for Government purposes notwithstanding any copyright notation hereon. We thank the National Science Foundation for support under grants DMR 0600073 and 0605890.


\end{document}